\documentstyle[12pt]{article}

  \newcommand{\be}{\begin{equation}}
  \newcommand{\ee}{\end{equation}}
  \newcommand{\bea}{\begin{eqnarray}}
  \newcommand{\eea}{\end{eqnarray}}
  \newcommand{\lb}{\label}
\renewcommand{\a}{\alpha}
  
\renewcommand{\b}{\beta}

\renewcommand{\d}{\delta}

  \newcommand{\mb}[1]{\mbox{\bf #1}}

  \newcommand{\ldel}{\langle}
  \newcommand{\rdel}{\rangle}
  \newcommand{\sprod}[2]{\ldel\,#1\,,\,#2\,\rdel}
  \newcommand{\lbrak}[2]{[#1,#2]}

  \newcommand{\bdm}{\begin{displaymath}}
  \newcommand{\edm}{\end{displaymath}}

  \newcommand{\we}{\wedge}

  \newcommand{\mtil}{\tilde{M}}
  \newcommand{\gtil}{\tilde{g}}
  
  \newcommand{\ghat}{\hat{g}}
  \newcommand{\cnull}{\chi_{(0)}}
  \newcommand{\ceins}{\chi_{(1)}}
  \newcommand{\cnullp}{\dot{\chi}_{(0)}}
  \newcommand{\ceinsp}{\dot{\chi}_{(1)}}
  \newcommand{\chip}{\dot{\chi}}
  \newcommand{\chipp}{\ddot{\chi}}
  \newcommand{\pip}{\dot{\pi}}
  \newcommand{\bnull}{B_{(0)}}
  \newcommand{\beins}{B_{(1)}}
  \newcommand{\pinull}{\pi_{(0)}}
  \newcommand{\pieins}{\pi_{(1)}}

  \newcommand{\tai}{\tau_r}
  \newcommand{\taii}{\tau_{\vartheta}}
  \newcommand{\taiii}{\tau_{\varphi}}

\begin{document}
\begin{flushright}
Z\"urich University Preprint\\
ZU-TH 9/95

\vskip 0.5\baselineskip
\end{flushright}

\begin{center}
\vfill
{\LARGE\bf Pulsations of spherically symmetric systems in general
relativity  }
\vfill
{\bf Othmar Brodbeck${}^1$, Markus Heusler${}^2$ and Norbert
Straumann${}^1$}
{${}^1$Institute for Theoretical Physics, University of
Z\"urich,\\ Winter\-thur\-er\-stras\-se\nobreak\ 190, 8057 Z\"urich,
Switzerland}\\
{${}^2$Enrico Fermi Institute, University of Chicago,\\
5640 S. Ellis Ave., Chicago, IL 60637}
\end{center}
\vfill
\begin{quote}
The pulsation equations for spherically symmetric black hole and
soliton solutions are brought into a standard form. The formulae apply
to a large class of field theoretical matter models and can easily
be worked out for specific examples. The close relation to the energy
principle in terms of the second variation of the Schwarzschild mass
is also established. The use of the general expressions is illustrated
for the Einstein-Yang-Mills and the Einstein-Skyrme system.
\end{quote}
\vfill
\newpage
\section{Introduction}
In recent years the stability properties of numerous new black hole and
soliton configurations, arising in selfgravitating nonlinear field
theories, have been studied. Early investigations of this kind were
concerned with the instabilities of such objects in SU$(2)$
Einstein-Yang-Mills theory. It was shown in \cite{NS,zhou} that both the
particle-like solutions of Bartnik and McKinnon \cite{bartnik}, as well
as the ``colored'' non-Abelian black holes \cite{volkov} are very fragile
structures. Some black holes with hair which were found in other matter
models turned out to be (at least linearly) stable. This is, for instance,
the case for the Skyrme model \cite{droz,heusler,heusler1} or when a Higgs
triplet is added to the Yang-Mills fields
\cite{lee,breitenlohner,aichelburg}.
If, however, a Higgs doublet is coupled, black holes and solitons turn out
to be unstable again \cite{boschung,winstanley}.

Most of these and other investigations are restricted to spherically
symmetric configurations. A closer look at them reveals many common
features. Hence, one would like to have a more general treatment,
encompassing a sufficiently large class of field theoretical models.
For any given one, it should then be possible, for example, to write
down readily the final pulsation equations, without repeating the
numerous intermediate computations.

A save and systematic method for a stability analysis is to linearize
the coupled field equations around a given equilibrium solution and to
study the frequency spectrum of the resulting (multicomponent)
eigenvalue problem. For spherically symmetric situations one expects
intuitively, that the resulting fluctuation operator is closely related
to the second variation of the Schwarzschild mass.
Similar ``energy principles''
are important in other fields of physics such as, for instance, in plasma
physics for ideal magnetohydrodynamic conditions \cite{chandrasekhar}.
In classical mechanics the second variation $\delta^2H$ of the Hamiltonian
provides the well-known Lagrange-Dirichlet stability criterion, which applies
usually also in infinite dimensions. (A counter example in elasticity
theory has been discovered by Ball and Marsden \cite{ball}.) One
should, however, remember that the energy criterion is often of no use.
A well-known example is provided by the equilibrium positions of the
restricted three-body problem. Here one has to linearize the Hamiltonian
vector field in order to investigate linear stability; $\delta^2H$ is
indefinite and gives no information on the fluctuation spectrum.

If the energy principle applies, it provides usually the simplest
method to decide on (linear) stability. In this paper we establish the
validity of the energy principle for a large class of matter models
coupled to gravity. More precisely: Under certain assumptions on the
dimensionally reduced form of the matter Lagrangian, we show that the
fluctuation operator of the pulsation equation can be read off from the
second variation $\delta^2 M$ of the Schwarzschild mass. Although this is
not surprising, we felt that a general demonstration of this is needed,
in particular since energy considerations in gravity are notoriously
problematic. More important, in deriving this connection we arrived at
explicit expressions for the pulsation equations and for $\delta^2 M$.
These can readily be worked out for a given field theoretical model,
which should considerably simplify future stability studies.

The article is organized as follows. In section 2 we specify the form
of the dimensionally reduced Lagrangian of the matter model. The
coupled field equations for (time-dependent) spherically symmetric
configurations are then expressed entirely in terms of this effective
Lagrangian, taken for a {\em diagonal\/} form of the metric on orbit
space (see eqs. (\ref{eq1}) -- (\ref{eq4})). This involves some
nontrivial considerations since, in the time-dependent case, one would
otherwise loose  one of Einstein's field equations. The general set up
is used in section 3 to bring the pulsation equations into a standard form,
involving only the matter perturbations. From these the metric
perturbations can be obtained algebraically, without solving any
differential equations. In section 4 we demonstrate that the
fluctuation operator can be read off from the second variation of the
total mass. The Einstein-Yang-Mills system and the Einstein-Skyrme
system are taken as illustrations for ready applications of our general
formulae. The dimensional reduction of the gravitational Lagrangian and
the derivation of the mass variation formula for spherically symmetric
configurations are deferred to two appendices.
\section{Dimensional Reduction}
\addtolength{\jot}{7pt}
\subsection{The effective Lagrangian}
\setcounter{equation}{0}
\renewcommand{\theequation}{\arabic{section}.\arabic{equation}}
Let us consider a spherically symmetric spacetime $(\mtil,\gtil)$. By
definition, $(\mtil,\gtil)$ is a Lorentz manifold on which SO$(3)$ acts
as an isometry group, such that all group orbits are metric
two-spheres. This guarantees that spacetime $(\mtil,\gtil)$ is locally
the warped product of a two-dimensional Lorentz manifold $(M,g)$
(the orbit space $\mtil/\mbox{SO}(3)=M$ with induced metric $g$) and
the two-sphere $S^2$ with standard metric $d\Omega^2$:
\be
\tilde M=M\times S^2,\qquad \gtil = g + r^2d\Omega^2\;.
\ee
The positive function $r\colon M\to\mbox{\rm R}$ (the
Schwarzschild coordinate) is assumed to have no critical points on $M$.
For the volume form $\tilde \eta$ on $\tilde M$ there is a
corresponding split: $\tilde\eta = \eta\wedge r^2 d\Omega$, where
$\eta$ and $d\Omega$ are the volume forms on $M$ and $S^2$,
respectively. In addition, we introduce the intrinsically defined
function $N=(dr|dr)$, where $(\,\cdot\;|\;\cdot \,)$ denotes the inner
product on $M$ and the mass fraction $m$ is defined by $N=1-2m/r$.

As is shown in appendix A, the dimensional reduction
of the Einstein-Hilbert action yields for the effective
gravitational Lagrangian $L_G$, after subtracting the standard
boundary term,
\be
L_G \, \eta \, = \,\frac{1}{N}\,(dm\,|\,dr) \, \eta \,.
\label{actgrav}
\ee
Expanding $dm$ with respect to a coordinate basis, $dm=\dot
m\,dt+m'\,dr$, and introducing the metric functions
$\beta=-(dt\,|\,dr)/N$ and $S=\sqrt{-g}$, the effective
Lagrangian takes the form
\be
L_G\,\eta\,=\,{\cal L}_G\,dt\we dr\nonumber\;,\qquad
{\cal L}_G\,=\,(m' - \beta \, \dot{m})S\;.
\label{actgrav1}
\ee
In terms of $N$, $S$ and $\beta$ the metric on $M$ becomes
\be
g = -NS^2 \, (dt^2
    \,+\,2\,\beta \, dt\,dr )
    \,+\,(\frac{1}{N}-\beta^2 NS^2)\,dr^2\;.
\lb{b1}
\ee

The effective Lagrangian ${\cal L}_G$ contains, as we shall show below,
the entire dynamical information. Requiring that the matter Lagrangian
depends only on the metric and its first derivatives, we shall now establish
the fact that the Einstein equations agree with the Euler-Lagrange equations
for the total effective Lagrangian.

Let us first consider the gravitational part. Defining the Euler-Lagrange
operator ${\rm\bf E}^0_f$ for a dynamical variable $f$, say, according to
\be
{\rm\bf E}^0_f \, = \, \Bigl\{\, \frac{\partial}{\partial f}
- \partial_r\Bigl(\frac{\partial}{\partial f'}\Bigr)
- \partial_t\Bigl(\frac{\partial}{\partial \dot{f}}\Bigr)\,
\Bigr\}\;\Big|_{\beta = 0}\quad,
\label{ELO}
\ee
we immediately find from eq. (\ref{actgrav1})
\be
{\rm\bf E}^0_m\, {\cal L}_G
 = - S'\;,\qquad
{\rm\bf E}^0_S\, {\cal L}_G
 = m'\;,\qquad
{\rm\bf E}^0_{\beta}\, {\cal L}_G
 =-\dot{m} S\;.
\ee
Using well known expressions for the
Einstein tensor $G_{\mu\nu}$ in Schwarzschild coordinates, we see that
\begin{eqnarray}
{\rm\bf E}^0_m\, {\cal L}_G
&=&-\frac{\,r\,}{2}SN^{-1} \, ({G^{\,r}}_{r} - {G^{\,t}}_{t}) \; ,
\label{ELO1}\\
{\rm\bf E}^0_S\, {\cal L}_G
&=& -\frac{\,r^2}{2} \, {G^{\,t}}_{t}\; ,
\label{ELO2}\\
{\rm\bf E}^0_{\beta}\, {\cal L}_G
&=& -\frac{\,r^2}{2}S \, {G^{\,r}}_{t} \; .
\label{ELO3}
\end{eqnarray}

Let us now turn to the effective matter Lagrangian ${\cal L}_M$,
defined by
\be
S_M\,=\,G^{-1}\int {\cal L}_M\;dt\wedge dr \;\;,
\label{LM}
\ee
where $S_M$ denotes the matter action. By virtue of the definition of
the energy momentum tensor $T_{\mu\nu}$, the variation of $S_M$ with
respect to the dynamical variable $f=N,S,\beta$ yields
\begin{eqnarray}
\d_f S_M \,
&=& \, \frac{1}{2}
\int T^{ab} \,\Bigl\{\frac{\partial g_{ab}}{\partial f}\,\d f
\Bigr\}\;\tilde\eta
\nonumber\\
&=& \,\frac{1}{2}
\int\,\Bigl\{ 4\pi r^2S\; T^{ab} \,\frac{\partial g_{ab}}{\partial f}
\,\Bigr\}
\d f\;dt\wedge dr \;\;.
\lb{b2}
\end{eqnarray}
If the matter Lagrangian depends on the metric and its first
derivatives only, this implies that ($\kappa=8\pi G$)
\be
{\rm\bf E}^0_{f}\, {\cal L}_M = \kappa\,\frac{r^2}{4}\,S\; T^{ab}
\,\frac{\partial g_{ab}}{\partial f}\Big |_{\beta=0}\;\;,\qquad
f=N,S,\beta\;.
\lb{b3}
\ee
Inserting  the parametrization (\ref{b1}) for the metric now yields the
desired result:
\begin{eqnarray}
{\rm\bf E}^0_m\, {\cal L}_M
&=&\kappa\,\frac{\,r\,}{2}SN^{-1} \, ({T^{\,r}}_{r} - {T^{\,t}}_{t}) \;
,
\label{b4}\\
{\rm\bf E}^0_S\, {\cal L}_M
&=&\kappa\,\frac{\,r^2}{2} \, {T^{\,t}}_{t}\; ,
\label{b5}\\
{\rm\bf E}^0_{\beta}\, {\cal L}_M
&=&\kappa\,\frac{\,r^2}{2}S \, {T^{\,r}}_{t} \; .
\label{b6}
\end{eqnarray}
These relations, together with eqs. (\ref{ELO1})\ --\ (\ref{ELO3}),
establish our assertion and enable us to write the Einstein equations
in the form
\be
m' = -{\rm\bf E}^0_S\, {\cal L}_M\; ,\qquad
S' = {\rm\bf E}^0_m\, {\cal L}_M\; ,\qquad
\dot{m}S = {\rm\bf E}^0_\beta\, {\cal L}_M\; .
\label{mSm}
\ee

The fact that $\beta$ may be set equal to zero after variation
reflects the freedom to diagonalize the metric of the orbit space $M$.
However, it is not surprising that (in a non-static situation)
one looses information by using a diagonal metric in the
effective action in the first place, that is,
before performing the variation.

Before we proceed, let us give two examples for which the above
reasoning applies, since, in these cases, the matter action contains no
derivatives of the metric.
\addtolength{\jot}{-7pt}
\addtolength{\jot}{7pt}
\subsection{Examples}
As a first example we consider the Einstein-Yang-Mills (EYM) system.
The matter action is
\be
S_{M}\,=\,\frac{1}{e^2}\,\mbox{Tr}\int\,F\wedge\tilde\ast F\;\;,
\ee
where $e$ is the (dimensionless) gauge coupling, $F = dA + A \we A$ is
the field strength assigned to the gauge potential $A$ and a star
$\tilde\ast$ denotes the Hodge dual with respect to the spacetime metric
$\tilde{g}$. For simplicity, we restrict ourselves to the gauge group
$SU(2)$. (The generalization to arbitrary gauge groups is straightforward
applying the equations derived in \cite{BRO}.) A spherically symmetric
gauge potential $A$ can then be represented as
\be
A \, = \,a\,\tai
\,+\,
(\mbox{\small Re}(w)-1)\,\Bigl\{\taiii\,d\vartheta
-\taii \sin\vartheta\,d\varphi\Bigr\}
\,+\,
 \mbox{\small Im}(w)   \,\Bigl\{\taii \,d\vartheta
 +\taiii\sin\vartheta\,d\varphi\Bigr\}\;,
\label{APHI}
\ee
where $a$ is a one-form on $M$, $w$ is a (complex) function on $M$ and
$\tai,\taii,\taiii$ denote the spherical generators of SU$(2)$, normalized
such that $\lbrak{\tai}{\taii}=\taiii$. A gauge transformation with
$U=\exp({\chi\tai})$, where $\chi$ is a function on $M$, preserves
the form of the potential $A$ and induces the transformations
\be
a\rightarrow a+d\chi\;,\qquad w\rightarrow \mbox{e}^{i\chi}\,w\;.
\label{indgaugetrans}
\ee
Hence, $a$ can be considered a gauge potential and $w$ a Higgs
field on $M$. With the parametrization (\ref{APHI}) for the potential
$A$ one finds for the matter action $S_{M}$,
\be
S_{M}\,=\,-\frac{4\pi}{e^2}\int\,
\biggl\{\, \frac{r^2}{2}(da|da)+(\,\overline{Dw}\,|\,Dw\,)+
\frac{(|w|^2-1)^2}{2r^2}
\,\biggr\}\,\eta \;,
\label{LYMH1}
\ee
where we have introduced the covariant derivative $Dw=dw-iaw$. Using
the parametrization (\ref{b1}) for the metric on $M$ and adopting the
temporal gauge $a=a_r\, dr$, we now easily obtain the following
expression for the effective matter Lagrangian ${\cal L}_{M}$:
\be
{\cal L}_{M} \, = \, {\cal L}_{M}^{(0)} + {\cal L}_{M}^{(\beta)}\, ,
\label{splitmatt}
\ee
with
\bea
\frac{1}{\a^2}{\cal L}_{M}^{(0)} & = &
\frac{1}{S}\,
\biggl\{\,\frac{1}{N}|\dot{w}|^2+\frac{r^2}{2}\dot{a}_r^2\,\biggr\}
\nonumber \\
&-&S \,\biggl\{ N \, \Big|w'-ia_rw\Big|^2 +
\frac{(|w|^2-1)^2}{2r^2}\,\biggr\}
\label{LM0}\;,\\
\frac{1}{\a^2}{\cal L}_{M}^{(\beta)} &=&
2 \,\beta N S \,\mbox{\small Re}\Bigl\{\dot{w}
(w'+ia_r\overline{w})\Bigr\} \label{LMbeta}
\eea
and $\a^2=4\pi G/e^2$. The Einstein equations are obtained from
eq. (\ref{mSm}). Applying the Euler-Lagrange operators ${\rm\bf
E}_{\overline{w}}$, ${\rm\bf E}_{a_r}$ on the diagonal part ${\cal
L}_{M}^{(0)}$ of the matter Lagrangian also yields the YM equations:
${\rm\bf E}_f{\cal L}_{M}^{(0)}={\rm\bf E}^0_f{\cal L}_{M}=0$ for
$f=\overline{w},a_r$. (In addition to these equations one has the
YM Gauss constraint, which got lost as a consequence of the gauge
fixing.)

As a second example we consider the Einstein-Skyrme model, for
which the matter action $S_M$ is (see \cite{heusler1} and references
therein)
\be
S_M \, = \, \frac{f^2}{4} \, \mbox{Tr}\int\,
A\wedge\tilde\ast A \;\, + \;\,
\frac{1}{16e^2} \, \mbox{Tr}\int\,
dA\wedge\tilde\ast dA \;.
\label{skyrme}
\ee
Here $f$ and $e$ are coupling constants, $A = U^{-1}dU$ and $U$ is a
$SU(2)$-valued function on $\tilde M$ (describing the pion field). The
``hedgehog ansatz''
\be
U \, = \, \cos \chi \, - \, 2\,\tau_r\sin \chi \; ,
\label{hedgehog}
\ee
with a function $\chi$ on the orbit space $M$, gives for the effective
matter Lagrangian ${\cal L}_M$,
\be
{\cal L}_{M} \, = \,{\cal L}_{M}^{(0)} + {\cal L}_{M}^{(\beta)}\, ,
\ee
where
\bea
\frac{1}{\a^2}{\cal L}_M^{(0)} &=&
\frac{1}{S} \,
( \frac{1}{N} \, u \, \dot{\chi}^2 )
\,-\,S \, (Nu \,  \chi'^2 +v) \;,
\label{l0s}\\
\frac{1}{\a^2}{\cal L}_M^{(\beta)} &=&
2\,\beta N S \, \, u \, \dot{\chi} \chi' \;,
\label{betaparts}
\eea
with $\a^2=2\pi G/e^2$ and
\be
u = (f e)^2 r^2 + 2 \sin^2 \chi\;,\qquad
v = (2(f e)^2 r^2 + \sin^2 \chi) \frac{\sin^2 \chi}{r^2}\;.
\ee
The entire set of field equations (including the Skyrme equation) can
be derived quickly from the total effective Lagrangian ${\cal
L}_G+{\cal L}_M$.

In the examples above, the diagonal part ${\cal L}_{M}^{(0)}$ of the
Lagrangian consists of a term proportional to $S^{-1}$ and a term
proportional to $S$, which, henceforth, will be called the ``kinetic''
and the ``potential'' parts, respectively. Note that the kinetic part
is quadratic in the time derivatives of the matter fields and that it
only contains terms proportional to $S^{-1}$ or ${(NS)}^{-1}$, whereas
the potential part consists only of terms proportional to $S$ or
$(NS)$.
\addtolength{\jot}{-7pt}
\addtolength{\jot}{7pt}
\subsection{Basic equations}
Motivated by the previous examples, we assume in the following that the
diagonal part ${\cal L}:={\cal L}_M^{(0)}$ of the effective matter
Lagrangian has the form
\be
{\cal L} \, = \,
\frac{1}{2S} \, \Bigl\{\,\dot\chi_{(0)}^2+\frac{1}{N}
\,\dot\chi_{(1)}^2  \,\Bigr\}
\,- \,
S \, (N \, U \, + \, P) \, ,
\label{call}
\ee
with {\/\em real\/} matter fields $\chi_{(0)}$ and $\chi_{(1)}$ and
positive definite quadratic forms $\dot\chi_{(0)}^2$ and
$\dot\chi_{(1)}^2$:
\be
\dot\chi_{(j)}^2=\sprod{\dot\chi_{(j)}}{B_{(j)}\dot\chi_{(j)}}\;,\qquad
j=0,1\;, \ee
where the inner products for the matter fields are  denoted by
$\sprod{\,\cdot\,}{\,\cdot\,}$. (For the EYM system we have, for
instance, $\cnull=a_r$, $\ceins=(\mbox{\small Re}\,w,\mbox{\small
Im}\,w)$ and $B_{(0)}=\a^2r^2$, $B_{(1)}=2\a^2\mbox{diag}(1,1)$.) We
further require (in view of the following sections) that the function
$P$ does not depend on spatial derivatives of the matter field
$\chi:=(\chi_{(0)},\chi_{(1)})$:
\[
B_{(j)}=B_{(j)}(\chi,\chi',;r)\;,\qquad
U=U(\chi,\chi';r)\;,\qquad
P=P(\chi;r)\;.
\]

With these assumptions for $\cal L$ it is possible to reconstruct the
total effective Lagrangian ${\cal L}_G+{\cal L}_M$ up to first order in
$\b$. We will show that
\be
{\cal L}_G+{\cal L}_M =  S (m'  - \beta\dot{m}) +
{\cal L}-\beta
\sprod{\dot{\chi}}{{\cal L}_{\chi'}}+O(\b^2) \; ,
\label{defl}
\ee
where $\chi:=(\chi_{(0)},\chi_{(1)})$ and ${\cal L}_{\chi'}:=\partial
{\cal L}/\partial \chi'$. For the field equations ${\rm\bf E}^0_f\,
\{{\cal L}_G+{\cal L}_M\}=0$, $f=m,S,\b,\chi$, this gives
\bea
m' &=& -{\cal L}_S \; ,
\label{eq1}\\
S' &=&\phantom{-}{\cal L}_m \; ,
\label{eq2}\\
\dot{m} &=& -S^{-1}\sprod{\dot{\chi}}{{\cal L}_{\chi'}}  \; ,
\label{eq3}\\
({\cal L}_{\dot{\chi}}) \dot{ } &=&
-({\cal L}_{\chi'})'
+{\cal L}_{\chi}\; ,
\label{eq4}
\eea
involving only the diagonal part ${\cal L}={\cal L}_M^{(0)}$ of the
effective matter Lagrangian.

To prove our claim, we show that the $\dot m$ equation (\ref{eq3})
follows from the remaining ones. We do so by rewriting the system in
Hamiltonian form. Defining the conjugate momenta
$\pi:=(\pi_{(0)},\pi_{(1)})$ by
\be
\pinull  =  {\cal L}_{\cnullp} \, = \, \frac{1}{S}\bnull\cnullp \; ,
\qquad\quad
\pieins  =  {\cal L}_{\ceinsp} \, = \, \frac{1}{NS}\beins\ceinsp \; ,
\label{momenta}
\ee
the effective matter Hamiltonian is obtained by
a Legendre transformation:
\bea
{\cal H} &=& \sprod{\pi}{\dot\chi} - {\cal L}\;,\nonumber\\
	 &=&\frac{S}{2}\Bigl\{{\pinull^2}+N\,{\pieins^2}\Bigr\}+
S(NU + P)\;,
\label{calh}
\eea
where $\pinull^2$, $\pieins^2$ denote the quadratic forms
\be
\pi_{(j)}^2=\sprod{\pi_{(j)}}{B_{(j)}^{-1}\,\pi_{(j)}}\;,\qquad
j=0,1\;.
\label{calAB}
\ee
Observing that
\be
{\cal L}_m = -{\cal H}_m\;,\qquad
{\cal L}_S=-{\cal H}_S= -S^{-1}{\cal H}
\ee
and
\be
\chip={\cal H}_{\pi}\;,\qquad
{\cal L}_{\chi} = -{\cal H}_{\chi}\;,\qquad
{\cal L}_{\chi'} = -{\cal H}_{\chi'}\;,
\ee
the field equations (\ref{eq1})\ --\ (\ref{eq4})
are equivalent to
\bea
m' & = & \phantom{-}{\cal H}_{S}\;,
\label{h1}\\
S'  & = & -{\cal H}_m\;,
\label{h1a}\\
\dot{m} & = & S^{-1}\sprod{{\cal H}_\pi}{{\cal H}_{\chi'}}  \; ,\\
\chip & = & \phantom{-}{\cal H}_{\pi} \;,\label{h2}\\
\pip  & = & ({\cal H}_{\chi'})'\,-\,{\cal H}_{\chi} \, .
\label{h2a}
\eea
Taking advantage of these formulae it is now rather simple
to complete the proof. Using the Einstein equations (\ref{h1}),
(\ref{h1a}) and,
subsequently, the matter equations (\ref{h2}), (\ref{h2a}), we obtain
\begin{eqnarray*}
(\dot{m} S)'
& = &
(S^{-1} {\cal H}) \dot{ } S \, - \, \dot{m} {\cal H}_m \\
& = &
\dot{\cal H} - \dot{S} {\cal H}_S - \dot{m} {\cal H}_m\\
& = &
\sprod{\pip}{{\cal H}_{\pi}} +
\sprod{\chip}{{\cal H}_{\chi}}+
\sprod{(\chi') \dot{ }}{{\cal H}_{\chi'}} \\
& = &
\sprod{\chip}{{\cal H}_{\chi'}}' \;,
\end{eqnarray*}
which is the spatial derivative of eq. (\ref{eq3}). (It would have
been considerably less
convenient to use the Lagrangian formulation
and it would have been even less so
to work with the matter equations for an explicitly given model.)

To summarize, the {\em diagonal\/} part (\ref{call}) of the matter
Lagrangian uniquely determines the field equations
(\ref{h1})\ --\ (\ref{h2a}). The last four of these equations form a
complete set of differential equations for $m,S,\pi,\chi$, and, as is
easily seen, imply that the constraint equation (\ref{h1}) (the
Hamilton constraint) propagates.

It will become clear in the following sections, that the above
setup is also well adapted for treating linear perturbations
of spherically symmetric selfgravitating configurations.
\addtolength{\jot}{-7pt}
\addtolength{\jot}{7pt}
\section{The Pulsation Equation}
\setcounter{equation}{0}
\renewcommand{\theequation}{\arabic{section}.\arabic{equation}}
In this section we linearize the field equations
(\ref{eq1})\ --\ (\ref{eq4}) in the vicinity of a static solution.
As we will see, the only metric perturbations which enter the first
order matter equations are the variations of $m$ and $\Delta r$, where
$\Delta $ denotes the Laplace operator for the orbit space $(M,g)$.
Using (\ref{eq1}) and (\ref{eq2}) one easily finds
\be
\Delta r =\frac{1}{S}(NS)'=
\frac{2m}{r^2}\,-\,\frac{1}{r}\,(\pinull^2 + 2P)\;.
\label{eq2bis}
\ee
Together with the linearized Einstein equations, these metric
perturbations can then be expressed in terms of the equilibrium
solution and the perturbations of the matter fields. This enables
one to derive a pulsation equation for the matter fields only.

In the following, $N,S,\chi$, etc.\ refer to a static solution
of (\ref{eq1})\ --\ (\ref{eq4}), whereas time-dependent perturbations
are denoted by $\d N,\d S,\d\chi$, etc.\ .

Let us first discuss the metric perturbations $\d m$ and $\d\Delta r$.
Since both $\dot{m}$ and $\dot{\chi}$ are quantities of first order,
the $\dot m$ equation (\ref{eq3}) yields
\be
(S\,\delta m)\dot{}
\,=\,-\sprod{{\cal L}_{\chi'}}{\delta \chi}\dot{}\;.
\ee
On the other hand, the Einstein equations
(\ref{eq1}) and (\ref{eq2}) imply that
\begin{eqnarray*}
(S \, \delta m)'
& = & {\cal L}_N \, \delta N \, - \,
S \, \delta {\cal L}_S\nonumber\\
& = &
S \, (N \delta U \, + \, \delta P)\;.
\end{eqnarray*}
Using the matter equation (\ref{eq4}) for the equilibrium
solution, we see that
\be
(S \, \delta m)'
= -\sprod{{\cal L}_{\chi'}}{\delta \chi}'\lb{dSdmdr}\;.
\ee
Hence,
\be
d\,(S\,\delta m) =
-\,d\,\sprod{{\cal L}_{\chi'}}{\delta \chi}\;.
\lb{ddm}
\ee
By assumption, $P$ does not depend
on derivatives of the matter fields. Equation (\ref{ddm}) thus yields
\be
\d m=  N\sprod{U_{\chi'}}{\delta \chi}\;.
\lb{delm}
\ee
Here the integration constant has been set equal to zero. In the
soliton case this is a consequence of the regularity requirement for
the center. For black holes, this is the correct choice since we
restrict our attention to variations with fixed position of the
horizon. (The relation between eq. (\ref{ddm})
and the first law for black holes is disussed in appendix B.)
The variation of eq. (\ref{eq2bis}) is now immediately found:
\bea
\delta \Delta r & = &
\frac{2}{r^2}\,\d m\,-\,\frac{2}{r}\,\d P\nonumber\\
& = &
  \frac{2N}{r^2}\,\sprod{U_{\chi'}}{\delta \chi}
\,-\, \frac{2}{r}\,\sprod{P_{\chi}}{\delta \chi} \;.
\label{deldelr}
\eea

Next, let us consider the matter equation (\ref{eq4}). Since, as will
be shown below, the metric perturbations can be eliminated from the
linearized matter equation, the latter can be written in the form
\be
\mbox{\bf T} \, \delta \chipp \, + \, \{\,\mbox{\bf U}_M\,+\, \mbox{\bf
U}_G\,\}\, \delta \chi \, = \, 0 \;,
\label{puls1}
\ee
where we have introduced the operators
\bea
\mbox{\bf T}\,\d\chi&=&NS\;\d\, ({\cal L}_{\dot{\chi}}) \dot{
}\;,\lb{bt}\\
\mbox{\bf U}_M\,\d\chi&=&NS\;\d\rlap{${}_\chi$}\;\;\Bigl\{({\cal
L}_{\chi'})'-{\cal L}_{\chi}\Bigr\}\;,\lb{bum}\\
\mbox{\bf U}_G\,\d\chi&=&NS\;\d\rlap{${}_g$}\;\;\Bigl\{({\cal
L}_{\chi'})'-{\cal L}_{\chi}\Bigr\}\;.\lb{bug}
\eea
Here, $\d_\chi$ and $\d_g$ denote the variations with respect to the
matter fields $\chi$ and the metric $g$, respectively. The first two
operators are immediately obtained from the definitions
(\ref{bt}) and (\ref{bum}):
\bea
\mbox{\bf T}\;\,
&=&
\left(
\begin{array}{cc}
N\bnull&0      \\
   0  &\beins\\
\end{array}
\right )\lb{Tmatrix}\>,\\
\mbox{\bf U}_M&=& \mb{p}_\ast\,U_{\chi'
\chi'}\,\mb{p}_\ast\,-\,i\,\lbrak{\,\mb{p}_\ast\,}{\,NS\,U_{\chi'\chi}\,}
\,+\,NS^2(NU_{\chi\chi}+P_{\chi\chi})\lb{UMmatrix}\;,
\eea
with the differential operator $\mb{p}_\ast=-iNS\,\partial/\partial r$.
For the operator $\mbox{\bf U}_G$ we find from eq. (\ref{bug})
\bea
\mbox{\bf U}_G\,\d\chi
&=&
NS^2\,\d_g\,\Bigl\{\frac{1}{S}\,\Bigl\{({\cal L}_{\chi'})'-{\cal
L}_{\chi}\Bigr\}\Bigr\}\;,
\nonumber\\
&=&
-NS^2\,\Bigl\{(U_{\chi'})'-U_\chi\Bigr\}\,\d
N\,-\,NS^2\,U_{\chi'}\,\d\Delta r\;.
\lb{bugdc}
\eea
The curly bracket in the last expression for $\mbox{\bf U}_G\,\d\chi$
can be further simplified:
\be
N\Bigl\{(U_{\chi'})'-U_\chi\Bigr\}=(N+2P-1)\,\frac{1}{r}\,U_{\chi'}+P_\chi\;,
\ee
as can be seen from eqs. (\ref{eq4}) and (\ref{eq2bis}) for
the unperturbed solution. Eventually we can use the expressions
(\ref{delm}) and (\ref{deldelr}) for $\d m$ and $\d \Delta r$ to
obtain the operator $\mbox{\bf U}_G$ from eq. (\ref{bugdc}),
\be
\mbox{\bf U}_G =
\frac{2NS^2}{r}
\biggl\{\,
U_{\chi'}\,\sprod{P_{\chi }}{\,\cdot\,}
\, + \,
P_{\chi }\,\sprod{U_{\chi'}}{\,\cdot\,}
\,+\,
(2P-1)\,\frac{1}{r}\,U_{\chi'}\,\sprod{U_{\chi'}}{\,\cdot\,}
\,\biggr\}\lb{UGmatrix}\;.
\ee
\addtolength{\jot}{-7pt}

Here we do not discuss mathematical properties, such as domains of
definition or (essential) self-adjointness for the operators in
eq. (\ref{puls1}). This was done by some of us for the EYM
system in a previous publication \cite{BRO1}.
%
%
%

\section{Mass Variation}
%
\setcounter{equation}{0}
\renewcommand{\theequation}{\arabic{section}.\arabic{equation}}
In this section we demonstrate that the second variation
of the total mass yields the same expression for the
fluctuation operator as the one previously derived by
means of linearizing the field equations. First of all,
we recall that the Komar formula provides one with an
expression for the total mass $M$ in terms of the
(asymptotically) timelike Killing field $k$. Defining the ``local" mass
$M(r)$ by the Komar expression over a
$2$-sphere with coordinate radius $r$ (\cite{heusler2}),
\be
M(r) \, = \, - \frac{1}{8 \pi}
\int_{S^2_r} \ast dk \, = \,
\frac{r^2}{2 S} (N S^2)' \, = \,
mS \, + \, r^2 N S' \, - \, r S m'  \, ,
\label{Komar}
\ee
one has $M = \lim_{\infty} M(r) = \lim_{\infty} m S$.
(Note that asymptotic flatness implies
$\lim_{\infty} r^2 S' = \lim_{\infty} r m' = 0$).
Thus, in a gauge where $\lim_{\infty} S(r) = 1$, we obtain
for configurations with $\delta \chi (\infty) = 0$
\be
\delta M \, = \, \lim_{\infty} S \delta m \, = \,
\lim_{\infty} S N \sprod{U_{\chi'}}{\delta \chi} \, = \, 0 \, ,
\lb{delm2}
\ee
where we have also used eq. (\ref{delm}) with $\d m(0)=0$ in the
soliton, and $\d m(r_H)=0$ in the black hole case.
Since $\delta M = 0$ for static solutions, we obtain the second order
formula
\be
E \, = \, M \, + \, \frac{1}{2} \, \delta^2 M \, = \,
M \, + \, \frac{1}{2} \int_{r_0}^{\infty} (S \, \delta^2 m)' \, dr \, ,
\label{E2order}
\ee
where $r_0=0$ for solitons and $r_0=r_H$ for black holes.

Our aim is to establish the relation
\be
\int_{r_0}^{\infty} (S \, \delta^2 m)' \, dr \, \, = \,
\int_{r_0}^{\infty} \Bigl\{ \,
\sprod{\delta \chip}{\mbox{\bf T} \delta \chip} \, + \,
\sprod{\delta \chi}{(\mbox{\bf U}_M + \mbox{\bf U}_G) \delta \chi} \,
\Bigr\} \, \frac{dr}{N S} \, ,
\label{secondM}
\ee
where the operators $\mbox{\bf T}$ and $\mbox{\bf U}_M + \mbox{\bf
U}_G$ in the kinetic and the potential parts of the fluctuation
operator are the same as in formula (\ref{puls1}). This is most
easily achieved by applying the Hamiltonian formulation presented
above. Using the Hamiltonian (\ref{calh}), the $m'$ equation
(\ref{h1}) becomes
\be
m' \, = \,
\frac{1}{2}\Bigl\{{\pinull^2}+N\,{\pieins^2}\Bigr\}+
NU + P \; .
\label{mprime}
\ee
Since $\chip$ is of first order, we have
$\delta \pinull = S^{-1} \bnull \delta \cnullp$ and
$\delta \pieins = (SN)^{-1} \beins$ $\delta \ceinsp$,
from which we obtain
\bdm
\delta^2 m' =
S^{-2} \sprod{\delta \cnullp}{ \bnull \delta \cnullp} +
S^{-2} \sprod{\delta \ceinsp}{N^{-1} \beins \delta \ceinsp} +
\delta^2 ( N  U + P) \, .
\edm
Recalling the definition (\ref{Tmatrix}) for $\mbox{\bf T}$
and using eq. (\ref{eq2}) for $S'$,
$S' = 2SU/r$, and
$\delta^2 m = -(r/2) \delta^2 N$,
this already yields the kinetic term in the expression
(\ref{secondM}) for the second variation of $m$,
\be
(S \delta^2 m)' \, = \, \frac{1}{N S}
\sprod{\delta \chip}{\mbox{\bf T} \delta \chip} \, + \,
S \, (N \delta^2 \, U \, + \, \delta^2 P) \, + \,
2S \, \delta N \, \delta U \, .
\label{V4}
\ee
Note that the terms involving the second variation of $N$ cancel, since
\[(S\d^2m)'=-SU\d^2N +S(\d^2m)'\;.\]

It remains to show that the integrals of the second and third term in
this
formula coincide with the corresponding expressions in eq.
(\ref{secondM}).
Using the definition (\ref{bum}) for $\mbox{\bf U}_M$,
the second term can be rewritten as follows:
\begin{eqnarray*}
S (N \delta^2 U + \delta^2 P)
& = & -\delta^2_{\chi} {\cal L} \\
& = & -\sprod{\delta_{\chi} {\cal L}_{\chi'}}{\delta \chi}'
 + \, \sprod{\delta_{\chi} \Bigl\{({\cal
L}_{\chi'})'-{\cal L}_{\chi}\Bigr\}}{\delta \chi}\\
& \doteq &
\frac{1}{N S} \, \sprod{\delta \chi}{\mbox{\bf U}_M \delta \chi} \, .
\end{eqnarray*}
Here and in the following `` $\doteq$ " stands for equal up to terms
whose spatial integration vanishes. (The integral over the last term
in the first line of the above formula does not contribute, since the
variations of the matter fields are required to vanish at $r = r_0$ and
for $r \rightarrow \infty$.) Hence, the integral of the second term in
eq. (\ref{V4}) assumes the desired form.

Our final task is to establish the equivalence between the last
terms in eqs. (\ref{V4}) and (\ref{secondM}). We do so by writing
$\delta U = U_{\chi} \delta \chi + U_{\chi'} \delta \chi'$
in  order to integrate $S \delta N \delta U$ by parts,
\bdm
S \delta N \delta U \, \doteq \,
\sprod{\delta \chi}{ \Bigl\{ U_{\chi} - (U_{\chi'}) \Bigr\} } \,
S \delta N \, - \,
\sprod{\delta \chi}{U_{\chi'}} \, (S \delta N)' \, .
\edm
The first term on the r.h.s. of this equation is already of the
desired form (cf. the first term in eq. (\ref{bugdc})). In order
to obtain the correct expression for the last term in eq. (\ref{V4}),
we have to add an additional $S \delta U \delta N$ to the above
formula.
Hence, it remains to compute
$S \delta U \delta N - \sprod{\delta \chi}{U_{\chi'}} (S \delta N)'$.
Using again $S' = 2SU/r$ and eq. (\ref{delm}) to express $\delta N$
in terms of $\delta \chi$, we find
\begin{eqnarray*}
S \delta U \delta N \, - \,
\sprod{\delta \chi}{U_{\chi'}} (S \delta N)'
&=& -  \sprod{\delta \chi}{U_{\chi'}} \,
\Bigl\{ N S \delta({2 U}/{r}) + (S \delta N)' \Bigr\}\\
&=&  -  \sprod{\delta \chi}{U_{\chi'}} \,
\Bigl\{ N S \delta({S'}/{S}) \, + \, (S \delta N)'  \Bigr\}\\
&=& -  \sprod{\delta \chi}{U_{\chi'}} \,
\delta \Bigl\{{(NS)'}/{S} \Bigr\} \, S \\
&=& - \sprod{\delta \chi}{U_{\chi'}} \,
\delta \Delta r \,S\; ,
\end{eqnarray*}
which is the desired expression, corresponding
to the second term in eq. (\ref{bugdc}).
Thus, the preceding two formulae together imply
\bea
2 \, S \delta U \delta N &\doteq&
S \, \sprod{\delta \chi}{\Bigl\{(U_{\chi'})'-U_\chi\Bigr\}\,\d
N\,-\,NS^2\,U_{\chi'}\,\d\Delta r} \,\nonumber\\
& = &
\frac{1}{NS}\sprod{\delta \chi}{\mbox{\bf U}_G \delta \chi}\;,
\eea
where we have also used the expression
(\ref{bugdc}) for the operator $\mbox{\bf U}_G$. This completes the
proof of the
variation formula (\ref{secondM}).

For the examples in section 2.2, with Lagrangians (\ref{LM0}) and
(\ref{l0s}), one can immediately read off the operators $\mbox{\bf T}$,
$\mbox{\bf U}_M$ and $\mbox{\bf U}_G$, given in (\ref{Tmatrix}),
(\ref{UMmatrix}) and (\ref{UGmatrix}), and thus write down the explicit
form of the pulsation equations (\ref{puls1}), as well as the second
variation of the energy (\ref{secondM}). The results and their
discussion can be found in \cite{volkov2} and \cite{droz}.
\section*{Acknowledgments}
We gratefully acknowledge financial support from the Swiss National
Science Foundation. One of us (MH) also wishes to thank the Enrico
Fermi Institute for its hospitality.
\vskip 3\baselineskip
\noindent{\Large\bf Appendix}
\appendix
\vskip \baselineskip
\section{The gravitational Lagrangian}
\setcounter{equation}{0}
\renewcommand{\theequation}{\Alph{section}\arabic{equation}}

This appendix is devoted to the dimensional reduction
of the spherically symmetric Einstein-Hilbert action.
Our aim is to derive the expression (\ref{actgrav}) for
the effective Lagrangian in terms of the intrinsically
defined function $r\colon M\to\mbox{\rm\bf R}$.
Recall that $M=\mtil/\mbox{SO}(3)$ is the orbit space
with induced metric $g$, and spacetime $(\mtil,\gtil)$ is
a warped product of $M$ and $S^2$ with metric $\gtil = g +  r^2\ghat$,
where $\ghat = d \Omega^2$. In what follows, we will use a tilde for
spacetime quantities and a hat for quantities on $S^2$.

A natural basis on the $2$-dimensional orbit space is
provided by the $1$-forms $dr$ and $\ast dr$, which
are used to introduce the orthonormal diade
$\{ \theta^0 , \theta^1 \}$,
\be
\theta^0 \, = \, - \ast \theta^1 \, ,
\; \; \; \;
\theta^1 \, = \, N^{-1/2} dr \, ,
\; \; \; \;
N \, = \, (dr | dr)\,,
\lb{A.2}
\ee
where $g = - \theta^0 \otimes \theta^0 + \theta^1 \otimes \theta^1$.
Here $\ast$ and $( \, \, | \, \, )$ denote the Hodge dual and the
inner product with respect to $g$, respectively.
Using $\Delta r \, \eta = \ast \triangle r = d \ast dr$,
we immediately find
\be
d \theta^0 \, = \, - \, \frac{1}{N} \, \Bigl\{ \,\frac{1}{2} dN -
\Delta r \, dr \,\Bigr\}
\, \wedge \theta^0 \, , \; \; \; \; \;
d \theta^1 \, = \, - \, \frac{1}{2N} \, dN \wedge \theta^1 \, .
\lb{A.3}
\ee
Comparing this with the structure equations and taking advantage of
$dN \wedge \theta^0 = (\ast dN) \wedge \theta^1$ and
$dN \wedge \theta^1 = (\ast dN) \wedge \theta^0$, enables
one to read off the expression for the connection form
$\omega = \omega^0_{\, \,  1} = \omega^1_{\, \,  0}$ of $M$,
\be
\omega \, = \, \frac{1}{N} \, \ast \,
( \, \frac{1}{2} dN \, - \, \Delta r \, dr \, ) \, .
\lb{A.4}
\ee
Using $N = 1 - 2m/r$, this also yields
\be
dr \we \omega \, = \,
\Bigl\{ \;\frac{m}{r^2} - \Delta r - \frac{(dm|dr)}{r N}\;\Bigr\} \,
\eta \, ,
\lb{A.5}
\ee
which will be used below.

The general expression for the Ricci scalar of a product
manifold with warping function $r$
(with norm $N = (dr|dr)$) is
\be
\tilde{R} \, = \, R \, + \, \frac{2}{r^2} \, (1 - N) \, - \,
\frac{4}{r} \, \Delta r \, .
\lb{A.6}
\ee
Since the second structure equation for a $2$-dimensional
pseudo-Riemannian manifold reduces to
$\Omega^0_{\; 1} = d \omega^0_{\; 1} = \frac{1}{2} R \eta$,
the Ricci scalar of $(M , g)$ is obtained from
\be
R \, \eta \, = \, 2 \, d \omega \, .
\lb{A.7}
\ee
Writing the volume form as
$\tilde{\eta} = \eta \wedge r^2 d \Omega$
we thus have
\be
\tilde{R} \tilde{\eta} \, = \,
2 \, \Bigl\{ \, r^2 d \omega \, + \, 2 \,
(\frac{m}{r} - r \Delta r) \, \eta \, \Bigr\} \,
\we \, d \Omega \, .
\lb{A.8}
\ee
Since we are interested in an effective Lagrangian involving no
second derivatives of the metric fields, we integrate the
first term by parts and subsequently use the
expression (\ref{A.5}) for $dr \we \omega$.
The Einstein-Hilbert action then becomes
\be
16 \pi \, S_G \, = \int_{\tilde{M}} \tilde{R} \, \tilde{\eta} \, = \,
2 \, \int_{\tilde{M}} d \, (r^2 \omega \we \d \Omega) \, + \,
4 \, \int_{\tilde{M}} \frac{(dm | dr)}{N} \, \eta \we d \Omega \, ,
\lb{A.9}
\ee
where the terms involving the Laplacian of $r$ have
canceled each other.

In order to obtain the correct effective Lagrangian after dimensional
reduction, it remains to subtract a boundary term $B$, involving the
trace of the extrinsic curvature, from the above formula.
In terms of the connection forms
$\tilde{\omega}_{\mu \nu}$ of spacetime, $B$ is given by (see, e.g.,
\cite{NS1})
\be
16 \pi \, B \, = \, \int_{\partial \tilde{M}} \tilde{\omega}_{\mu \nu}
\we
\tilde{\ast}  (\theta^{\mu} \we \theta^{\nu}) \, ,
\lb{A.10}
\ee
where $\partial \tilde{M} = \partial M \times S^2$ and it is assumed
that
the induced metric on $\partial M$ is kept fixed.
Using the connection forms
\be
\tilde{\omega}^a_{\, \, b} = \omega^a_{\, \, b} \, ,
\; \; \; \; \;
\tilde{\omega}^A_{\, \, B} = \hat{\omega}^A_{\, \, B} \, ,
\; \; \; \; \;
\tilde{\omega}^A_{\, \, b} = \frac{r,_b}{r} \theta^A \, ,
\lb{A.11}
\ee
and the expressions
$\tilde{\ast} (\theta^a \wedge \theta^b) = - \epsilon_{ab} r^2 d
\Omega^2$,
$\tilde{\ast} (\theta^A \wedge \theta^B) = \epsilon_{AB} \eta$ and
$\tilde{\ast} (\theta^a \wedge \theta^B) = - \ast \theta^a \wedge
\hat{\ast} \theta^{B}$,
we obtain
\be
\tilde{\omega}_{\mu \nu} \we
\ast (\theta^{\mu} \we \theta^{\nu}) \, = \,
2 \, r^2 \omega \, \wedge d \Omega^2 \, - \,
4 r \ast dr \, \wedge d \Omega^2 \, + \,
2 \hat{\omega}_{23} \, \wedge \eta \, .
\lb{A.12}
\ee
The last $3$-form in the above expression
does not contribute to the boundary integral,
since it contains no volume-form $d \Omega$.
In addition, the second term needs not be taken into account
neither, since we do not consider variations with respect to $r$. (If
one allows for variations with respect to $r$, this term contributes to
the effective Lagrangian. As a matter of fact, in this way one obtains
an effective
Lagrangian which, in addition to the $(ab)$-components, yields also
the angular components of the Einstein tensor.)

Hence, considering $r$ as a coordinate,
only the first term in eq. (\ref{A.12}) contributes
under variations,
\be
16 \pi \, \delta \, B \, = \, 2 \, \delta \,  \int_{\partial \tilde{M}}
r^2 \,
\omega \, \wedge \, d \Omega \, .
\lb{A.13}
\ee
Applying Stokes' theorem, we observe that this term cancels
the first term in the variation of the
dimensionally reduced Ein\-stein-Hil\-bert action (\ref{A.9}).
We thus obtain the desired result
\be
16 \pi \, \delta \, (S \, - \, B) \,
= \, 4  \delta \, \int_{\tilde{M}} \frac{(dm | dr)}{(dr | dr)} \,
\eta \wedge d \Omega \, = \, 16 \pi \, \delta \, \int_{M}
\frac{(dm | dr)}{N} \,
\eta \, ,
\lb{A.14}
\ee
completing the derivation of eq. (\ref{actgrav}).
%
%
%
\section{The First Law}
\setcounter{equation}{0}
\renewcommand{\theequation}{\Alph{section}\arabic{equation}}

In this appendix we establish that the
first law for spherically symmetric, static black
hole configurations reads
\be
\delta M \, - \, \frac{1}{8 \pi} \kappa \delta A \,
= \, \int_{r_H}^{\infty} (S \, \delta m)' \, dr \, ,
\lb{B.1}
\ee
where $\kappa$ and ${\cal A}$ denote the surface
gravity and the area of the horizon, respectively.
This expression is valid for arbitrary matter models.
For the theories under consideration in this article
the term on the r.h.s. generically does not contribute. This is
due to the fact that the integral is, by (\ref{delm}), equal to
$\lim_\infty [S N \sprod{U_{\chi'}}{\delta \chi}]$, which usually
vanishes as a consequence of asymptotic flatness.
The above formula can, of course, be obtained from
evaluating the general mass variation formula \cite{carter}
\be
\delta M - \frac{1}{8 \pi} \kappa \delta A =
\frac{1}{16 \pi} \int_{\Sigma} G^{\mu \nu}
\delta g_{\mu \nu} \ast k - \frac{1}{8 \pi}
\delta \int_{\Sigma} \ast G(k)
\lb{B.2}
\ee
(with $G(k)_{\mu} = G_{\mu \nu} k^{\nu}$)
in the spherically symmetric metric used throughout
this article. Here we shall, however, present a direct
derivation of eq. (\ref{B.1}), which is adapted to the spherical
symmetry.

Using the expression (\ref{Komar}) for the ``local''
Komar mass and the requirement of asymptotic flatness,
$M = \lim_{\infty} M(r) = \lim_{\infty} (m S)$, we have
\be
\delta M = \lim_{\infty}(S \delta m) \, = \,
\int_{r_H}^{\infty} (S \delta m)' \, dr \, + \,
(S \, \delta m)(r_H) \, ,
\lb{B.3}
\ee
where we have also used $\lim_{\infty}S = 1$ and $\lim_{\infty} \delta
m = \delta \lim_{\infty} m$.
Hence, in order to establish the desired result, we have to
show that
\be
\frac{1}{8 \pi} \kappa \delta A \, = \, (S \, \delta m)(r_H) \, .
\lb{B.4}
\ee
To see this, we first note that $2m(r_H)=r_H$ yields
\be
(\delta m)(r_H) \, = \, \frac{1}{2} (1 - 2 m'(r_H)) \, \delta r_H \, .
\lb{B.5}
\ee
In order to complete the derivation, we recall the general
result $M_H = \frac{1}{4 \pi} \kappa A$ for the evaluation of
the Komar integral over the Horizon. Comparing this with the
expression $M_H=M(r_H) = [S (m - rm')](r_H)$ which is obtained from
eq.\ (\ref{Komar}), the surface gravity of the horizon becomes
\be
\kappa \, = \frac{S_H}{2r_H}(1 - 2 m'(r_H)) \, .
\lb{B.6}
\ee
Using this  and $\delta A = 8 \pi r_H \delta r_H$
in eq.\ (\ref{B.5}) yields the desired result (\ref{B.4}).
Together with eq.\ (\ref{B.3}), this eventually establishes the
variation formula (\ref{B.1}).

\end{document}